\documentstyle[aps,epsf,rotate,multicol]{revtex}

\begin{document}

\draft

\title{Ivory Tower Universities and Competitive Business Firms }

\author{Vasiliki Plerou$^{1,2}$, Lu\'{\i}s A.\ Nunes
Amaral$^{1}$, Parameswaran Gopikrishnan$^{1}$,\\
Martin Meyer$^{1}$, and H.\ Eugene Stanley$^{1}$}

\address{$^{1}$ Center for Polymer Studies and Dept. of Physics,
        Boston University, Boston, MA 02215, USA \\
         $^{2}$ Dept. of Physics, Boston College, 
        Chestnut Hill, MA 02167, USA }

\date{Last Modified: June 11,1999 : Printed \today}

\maketitle

\begin{abstract}

There is nowadays considerable interest on ways to quantify the
dynamics of research activities, in part due to recent changes in
research and development (R\&D)
funding\cite{Moed97,Miflin97,Buffetaut98,Anderson94,Holden97,Halpern97,
Geissman97,Gazzaniga98,Cooper98}.  Here, we seek to quantify and
analyze university research activities, and compare their growth
dynamics with those of business
firms\cite{Gibrat31,Simon77,Sutton97,Stanley96,Takayasu98}.
Specifically, we analyze five distinct databases, the largest of which
is a National Science Foundation database of the R\&D expenditures for
science and engineering of 719 United States (US) universities for the
17-year period 1979--1995.  We find that the distribution of growth
rates displays a ``universal'' form that does not depend on the size
of the university or on the measure of size used, and that the width
of this distribution decays with size as a power law. Our findings are
quantitatively similar to those independently uncovered for business
firms\cite{Gibrat31,Simon77,Sutton97,Stanley96,Takayasu98}, and
consistent with the hypothesis that the growth dynamics of complex
organizations may be governed by universal mechanisms. 

\end{abstract}

\begin{multicols}{2}

  In the study of physical systems, the scaling properties of
fluctuations in the output of a system often yield information
regarding the underlying processes responsible for the observed
macroscopic behaviour\cite{Vicsek92,Barabasi95}.  Here, we analyze the
fluctuations in the growth rates of university research activities,
using five different measures of research activity.  The first measure
of the size of a university's research activities that we consider is
R\&D expenditures.  The rationale for using R\&D expenditures as a
measure of research activity is that research is an expensive activity
that the university finances with external support.

\begin{figure}
\narrowtext
\centerline{
\epsfysize=0.75\columnwidth{\rotate[r]{\epsfbox{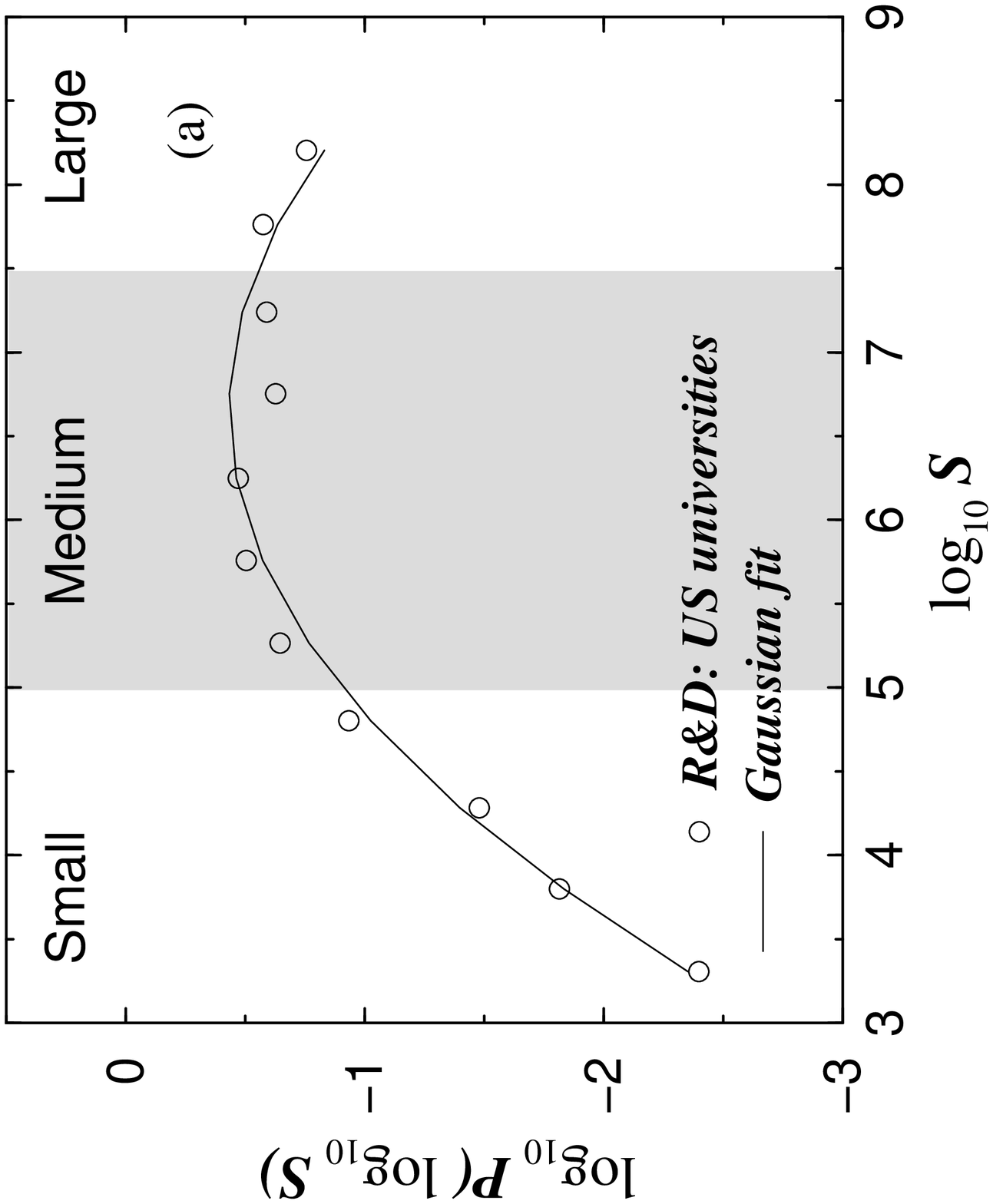}}}
}
\vspace*{0.5cm}
\centerline{
\epsfysize=0.75\columnwidth{\rotate[r]{\epsfbox{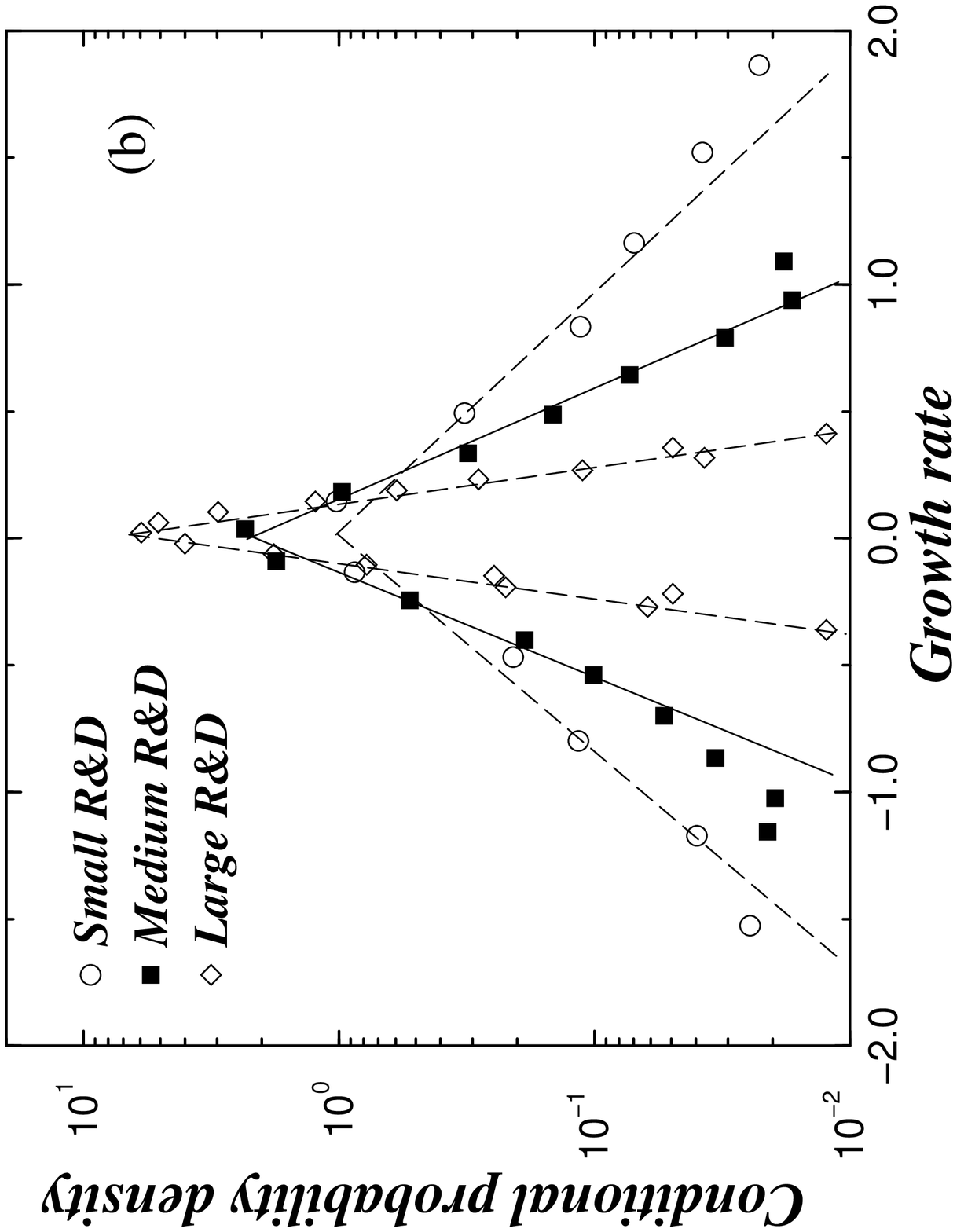}}}
}
\vspace*{0.5cm}
\centerline{
\epsfysize=0.75\columnwidth{\rotate[r]{\epsfbox{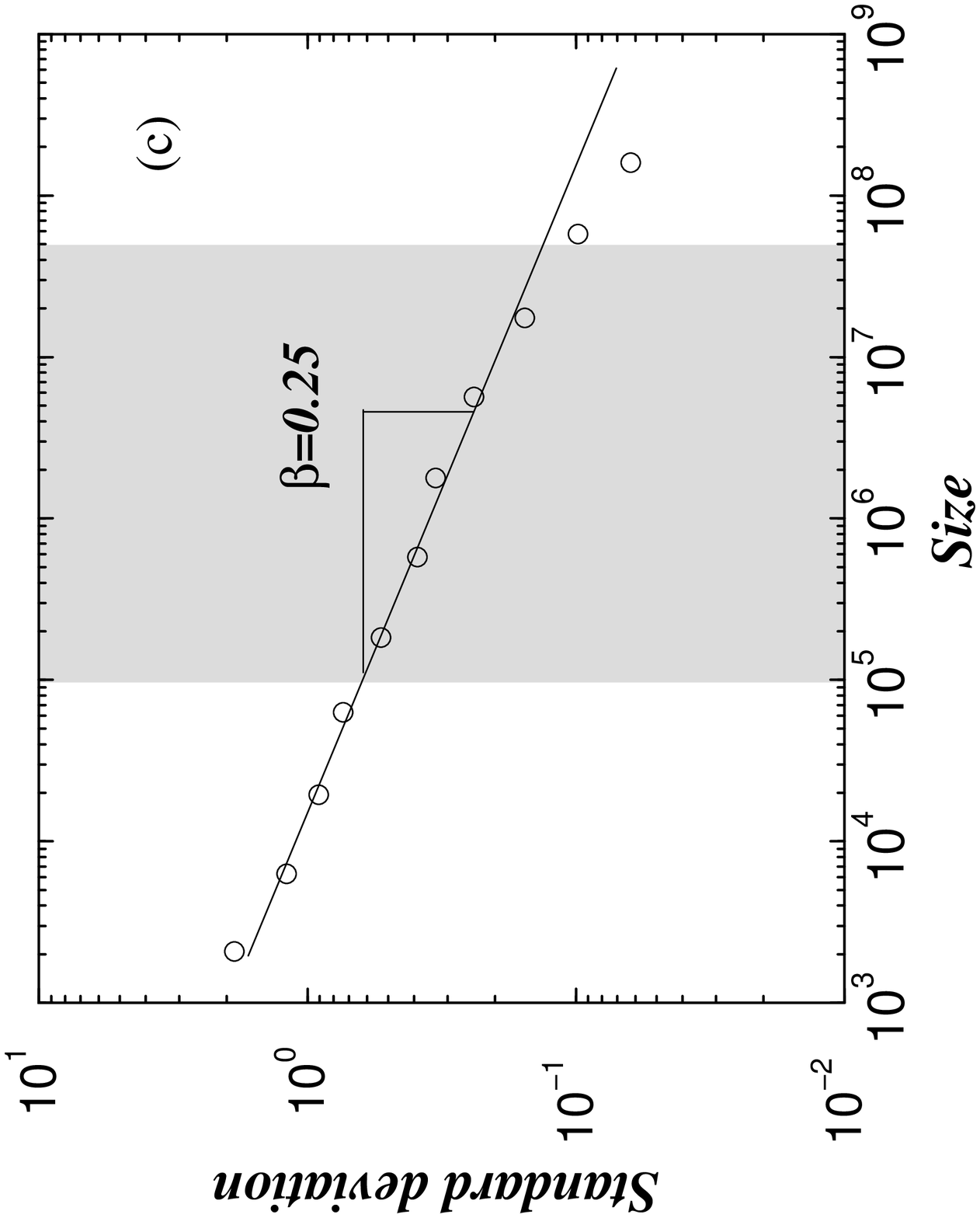}}}
}
\vspace*{0.5cm}
\centerline{
\epsfysize=0.75\columnwidth{\rotate[r]{\epsfbox{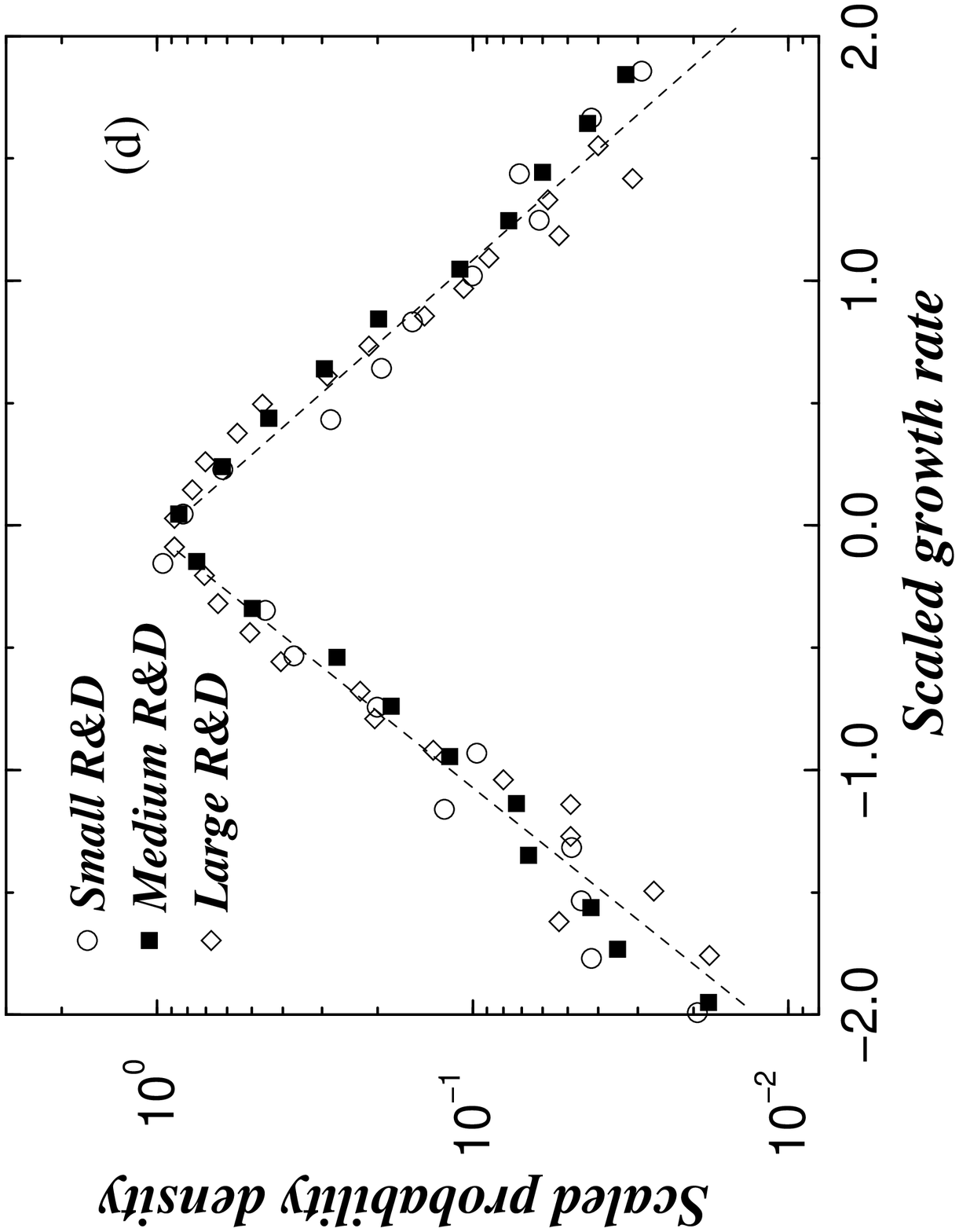}}}
}
\caption{ Growth dynamics of research activities at universities. {\it
a)} Histogram of the logarithm of the annual R\&D expenditures of 719
US universities for the 17-year period 1979--1995, expressed in 1992
US dollars. Here, $S$ denotes the R\&D expenditures detrended by
inflation so that values for different years are comparable.  The bins
were chosen equally spaced on a logarithmic scale with bin size
0.5. The line is a Gaussian fit to the data, which is a prediction of
Gibrat's theory\protect\cite{Gibrat31,Sutton97}.  {\it b)} Conditional
probability density function $p(g|S)$ of the annual growth rates $g$.
For this plot the entire database is divided into three groups
(depicted in (a) by different shades).  {\it c)} Standard deviation
$\sigma(S)$ of the distribution of annual growth rates of as a
function of $S$. The straight line is a power law fit to the the data,
and its slope gives the exponent $\beta=0.25\pm0.05$.  {\it d)} Scaled
probability density function $p(g|S) / \sigma^{-1}(S)$ plotted against
the scaled annual growth rate $(g - \bar g)/ \sigma(S)$ for the three
groups defined in (b). Note that the scaled data collapse onto a
single curve. }
\label{f.empirical}
\end{figure}

  We first analyze a database containing the annual R\&D expenditures
for science and engineering of 719 US universities\cite{NSF_RD} for
the 17-year period 1979--1995 ($\approx$ 12,000 data points).  The
expenditures are broken down by school and department.  The annual
growth rate of R\&D expenditures is, by definition, $g(t) \equiv \log[
S(t+1) / S(t)]$, where $S(t)$ and $S(t+1)$ are the R\&D expenditures
of a given university in the years $t$ and $t+1$ respectively.  We
expect that the statistical properties of the growth rate $g$ depend
on $S$, since it is natural that the fluctuations in $g$ will decrease
with $S$.  Therefore, we partition the universities into groups
according to the size of their R\&D expenditures
(Fig.~\ref{f.empirical}a).  Figure~\ref{f.empirical}b suggests that
the {\it conditional\/} probability density, $p(g | S)$, has the same
functional form, with different widths, for all $S$.

  We next calculate the width $\sigma(S)$ of the distribution of
growth rates as a function of $S$.  Figure~\ref{f.empirical}c shows
that $\sigma(S)$ scales as a power law
\begin{equation}
\sigma(S) \sim S^{-\beta},
\label{e-sigma}
\end{equation}
with $\beta = 0.25 \pm 0.05$. In Fig.~\ref{f.empirical}d, we
collapse the {\it scaled} conditional probability distributions onto a
single curve.

\begin{figure}
\narrowtext
\centerline{
\epsfysize=0.8\columnwidth{\rotate[r]{\epsfbox{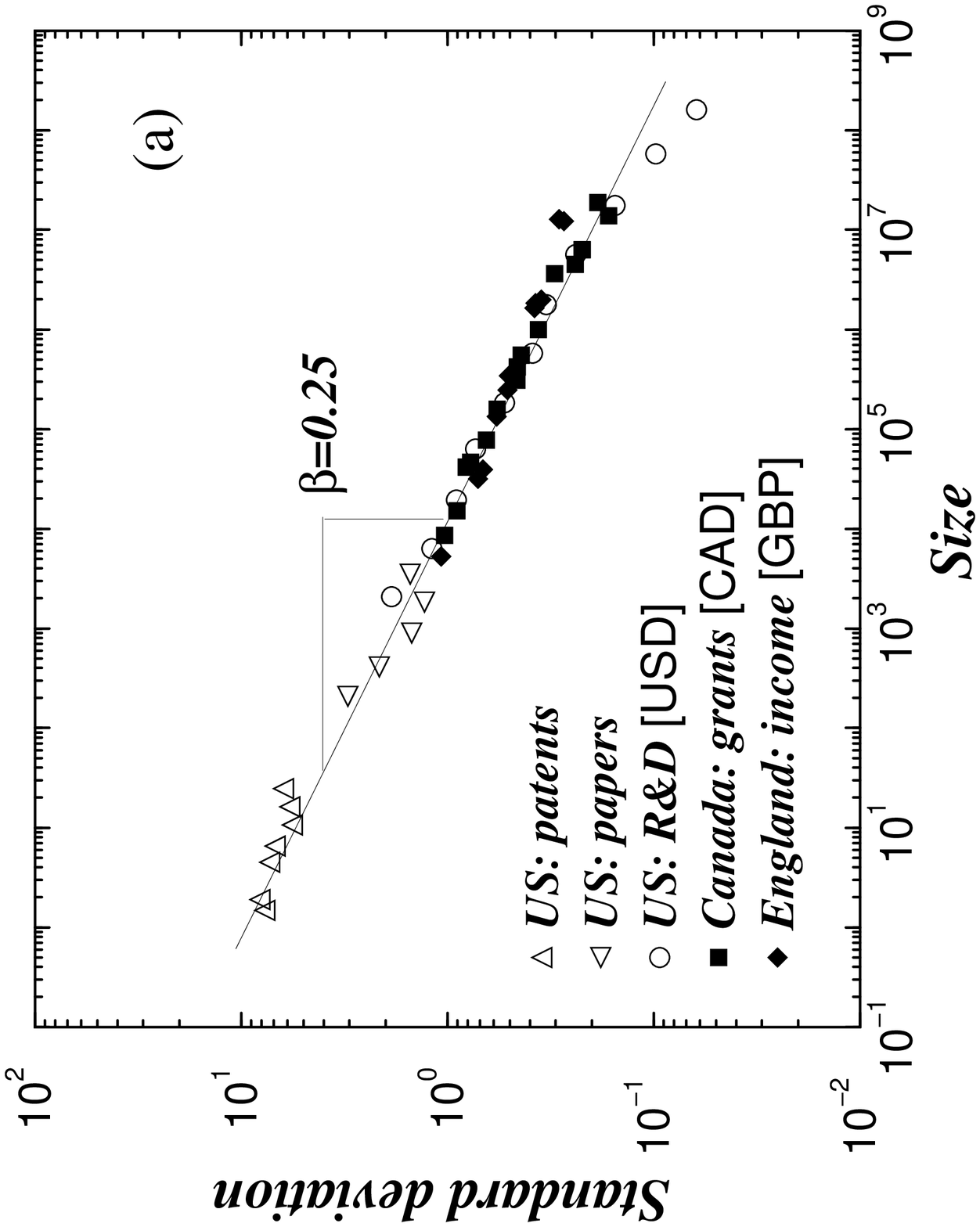}}}
}
\vspace*{0.5cm}
\centerline{
\epsfysize=0.8\columnwidth{\rotate[r]{\epsfbox{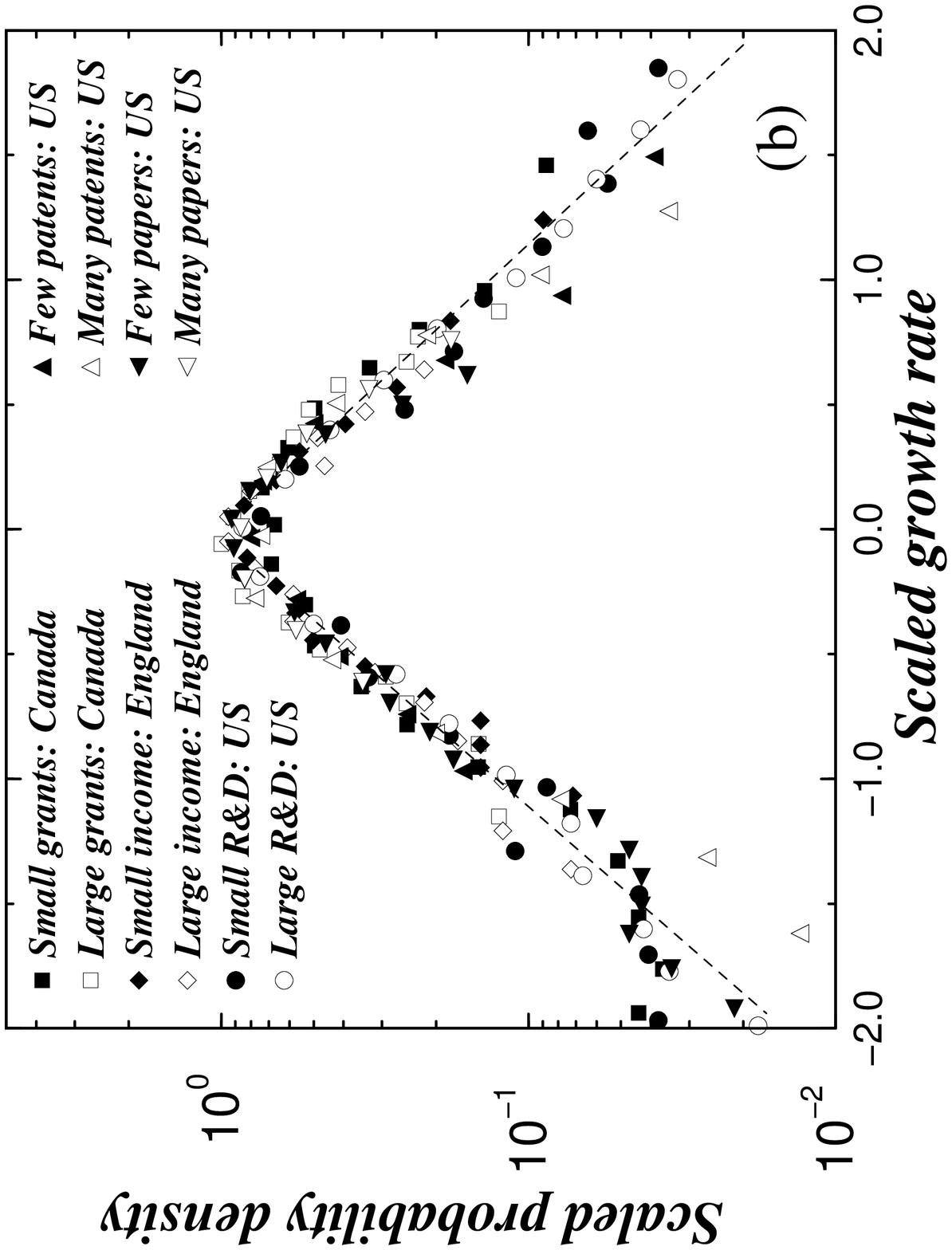}}}
}
\centerline{
\epsfysize=0.8\columnwidth{\rotate[r]{\epsfbox{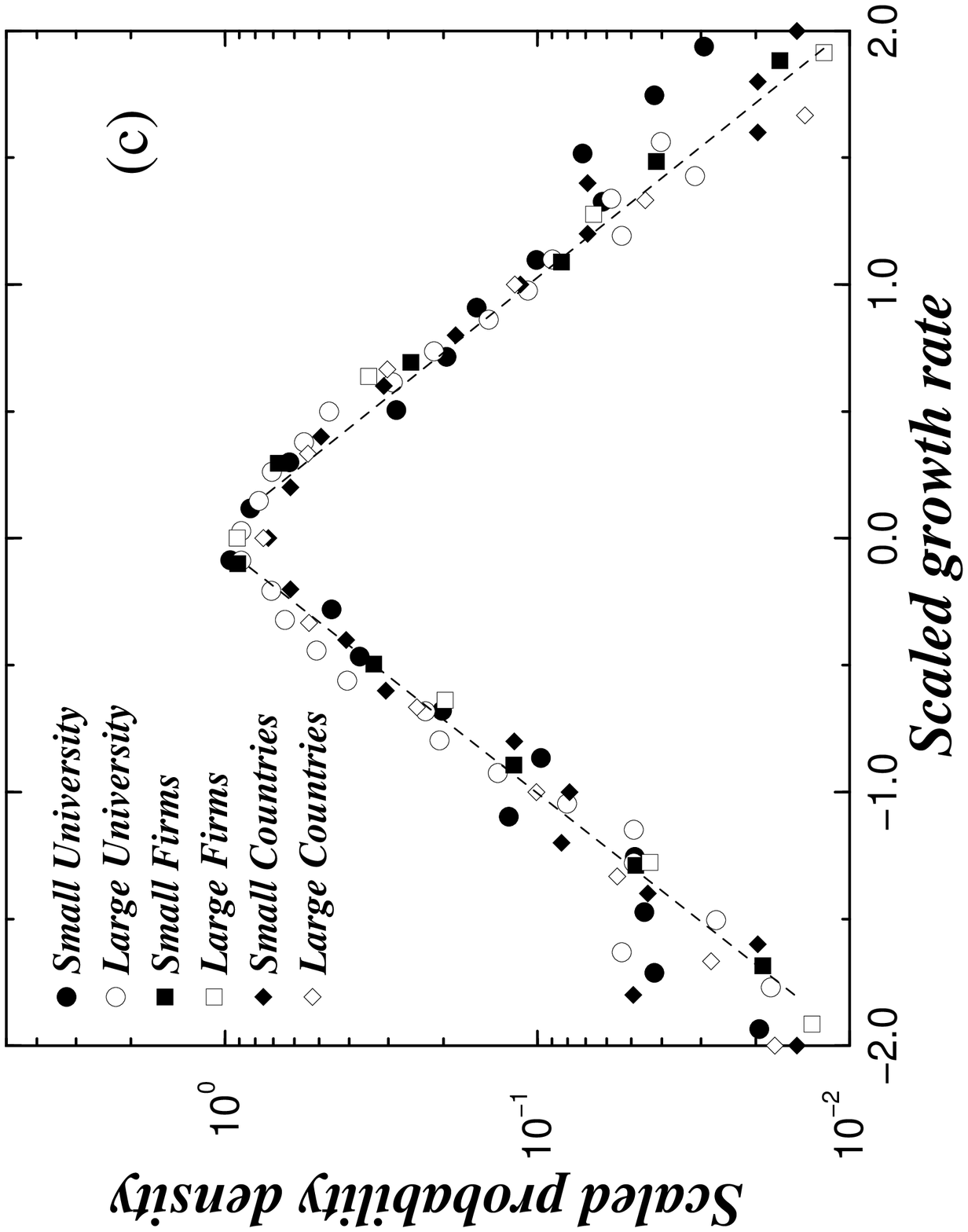}}}
}
\vspace*{0.5cm}
\caption{ Robustness of empirical findings for the distribution of
growth rates. {\it a)} Standard deviation $\sigma(S)$ of the
distribution of annual growth rates for different measures of research
activities and different academic systems from the data in the five
distinct databases analyzed: (i) the number of papers published each
year at 112 US universities, (ii) the number of patents issued each
year to 106 US universities, (iii) the R\&D expenditures in US dollars
of 719 US universities, (iv) the total amount in Canadian dollars of
the grants to 60 Canadian universities, and (v) the external incomes
in British pounds of 90 English universities.  It is apparent that for
all measures and all academic systems analyzed, we find a power law
dependence ---with the same exponent $\beta \approx 1/4$.  The values
of $\sigma$ for the different measures were shifted vertically for
better comparison of the estimates of the exponents.  {\it b)} The
distribution of annual growth rates, scaled as in
Fig.~\protect\ref{f.empirical}d, for the five databases. We show the
distribution of growth rates for 2 different groups, obtained in a way
similar to that described in Fig.~\protect\ref{f.empirical}b, for each
of the five measures. The data appear to collapse onto a single curve,
suggesting that the different measures have similar statistical
properties. {\it c)} The distribution of scaled annual growth rates
for different organizations: R\&D expenditures of US universities,
sales of firms, and GDP of countries.  The data collapse onto a single
curve suggesting that the scaled distributions have the same
functional form.  }
\label{f.collapse}
\end{figure}

  To test if these results for the dynamics of R\&D expenditures are
valid for other measures of research activity, we next analyze another
measure of a university's research activities, the number of papers
published each year\cite{Braun96,Lewison98,Schwarz98}.  We analyze
data for the 17-year period 1981--1997 from the {\it US University
Science Indicators\/} \cite{ISI}, which records the number of papers
published by the top 112 US universities ($\approx$ 1,900 data
points).  We find that the analog of Fig.~\ref{f.empirical} holds.
Particularly striking is the fact that the same exponent value, $\beta
= 1/4$, is found (Fig.~\ref{f.collapse}a) and that the same functional
form of $p(g | S)$ is displayed (Fig.~\ref{f.collapse}b).

  Next, we consider as a measure of size the number of patents issued
to a university\cite{Narin95}.  We ``manually'' retrieve from the
webpages of the {\it US Patent and Trademark Office\/}'s
database\cite{USPTO} the number of patents issued to each of 106
universities each year of the 22-year period 1976--1997 ($\approx$
2,300 data points).  We confirm that the analog of
Fig.~\ref{f.empirical} holds, with the same exponent value, $\beta =
1/4$ (Fig.~\ref{f.collapse}a), and the same functional form of $p(g |
S)$, Fig.~\ref{f.collapse}b.

  To test if our findings hold for different academic systems, we
analyze two databases on research funding of English\cite{HEFCE} and
Canadian\cite{NSERC} universities.  The same quantitative behavior is
found for the distribution of growth rates and for the scaling of
$\sigma$, with the same exponent value (Fig.~\ref{f.collapse}a) and
the same functional form of $p(g | S)$, Fig.~\ref{f.collapse}b.  Thus,
the analysis of all five databases confirms that the same quantitative
results hold across different measures of research activity and
academic systems.


  We next address the question of how to interpret our empirical
results.  We start with the observation that research is an expensive
activity, and that the university must ``offer'' its research to
external sources such as governmental agencies and business firms.
Thus, an increase in R\&D expenditures at university $A$ and a
decrease at university $B$ implies that the funders of research
increasingly choose their research from university $A$ as opposed to
university $B$\cite{Moed97}.  This qualitative picture parallels the
competition among different business firms, so it is natural to
enquire if there is quantitative support for this analogy between
university research and business activities.  To quantitively test
this analogy, we note that the results of Fig.~\ref{f.empirical} are
remarkably similar to the results found for
firms\cite{Stanley96,Takayasu98} and
countries\cite{Summers91,Durlauf96,Lee98}.  We plot in
Fig.~\ref{f.collapse}c the scaled conditional probabilities $p(g|S)$
for countries, firms and universities, and find that the distributions
for the different organizations fall onto a single curve.

  There is, however, one difference: For firms and countries, we find
$\beta \approx 1/6$, while for universities, $\beta \approx 1/4$.  We
can understand this difference using a model for organization
growth\cite{Amaral98}. In the model, each organization ---university,
firm, or country--- is made up of units.  The model assumes these
units grow through an independent, Gaussian-distributed, random
multiplicative process with variance ${\cal W}^2$.  Units are absorbed
when they become smaller than a ``minimum size'', which is a function
of the activity they perform. Units can also give rise to new units if
they grow by more than the minimum size for a new unit to form.  The
model predicts $\beta = {\cal W} / [2({\cal W} +{\cal D})]$, where
${\cal D}$ is the width of the distribution of minimum sizes for the
units\cite{Amaral98}.  For firms, the range of typical sizes is very
broad ---from small software and accounting firms to large oil and
automobile firms--- suggesting a large value of ${\cal D}$.  On the
other hand, for universities, the range of typical sizes is much
narrower, suggesting a small value of ${\cal D}$ and implying a larger
value of $\beta$ than for business firms.  This is indeed what we
observe empirically.

\begin{figure}
\narrowtext
\centerline{
\epsfysize=0.8\columnwidth{\rotate[r]{\epsfbox{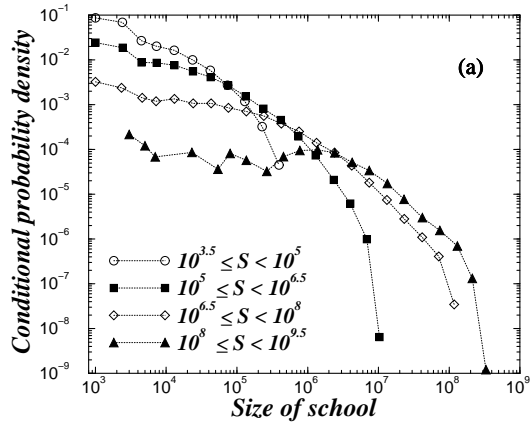}}}
}
\vspace*{0.5cm}
\centerline{
\epsfysize=0.8\columnwidth{\rotate[r]{\epsfbox{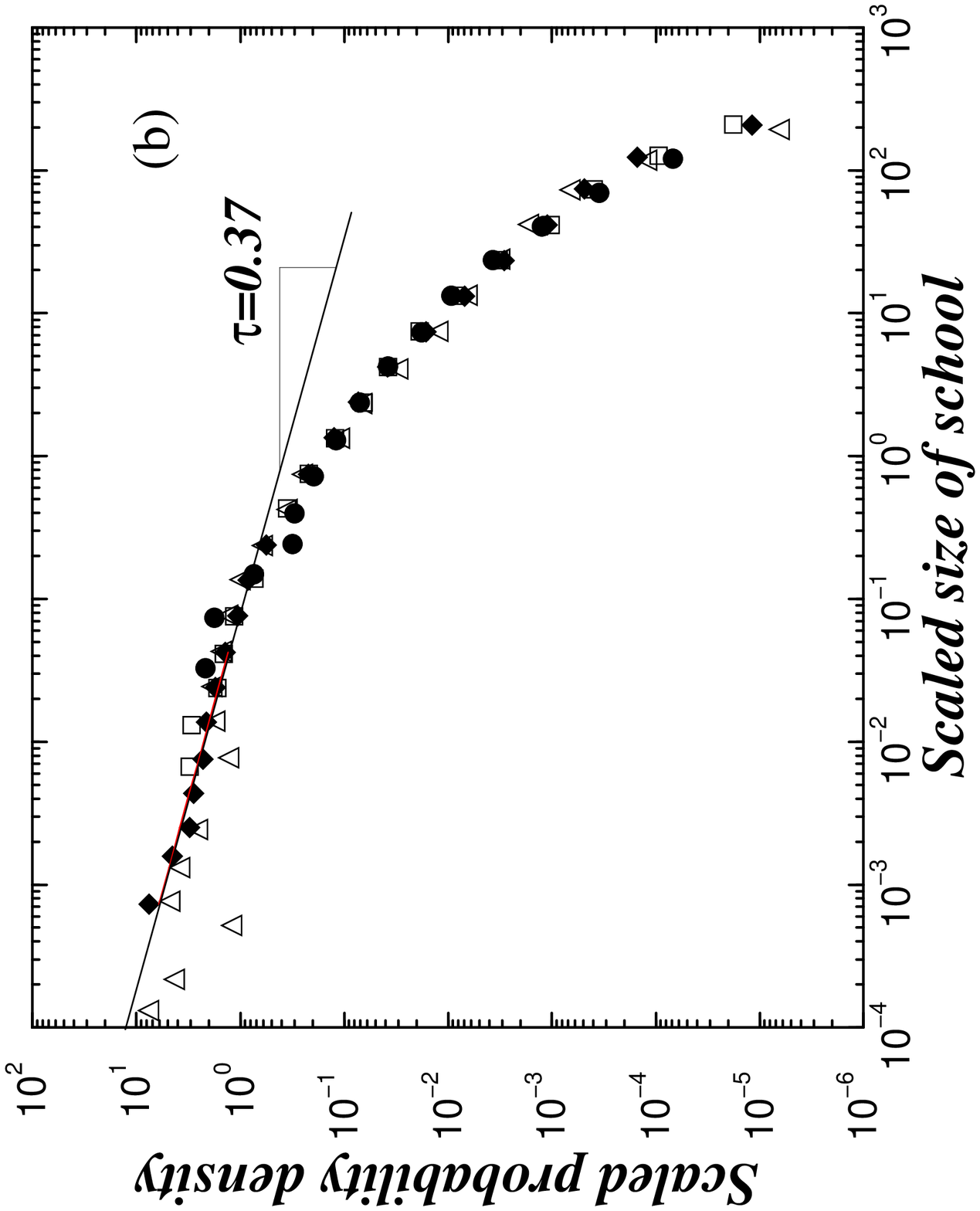}}}
}
\centerline{
\epsfysize=0.8\columnwidth{\rotate[r]{\epsfbox{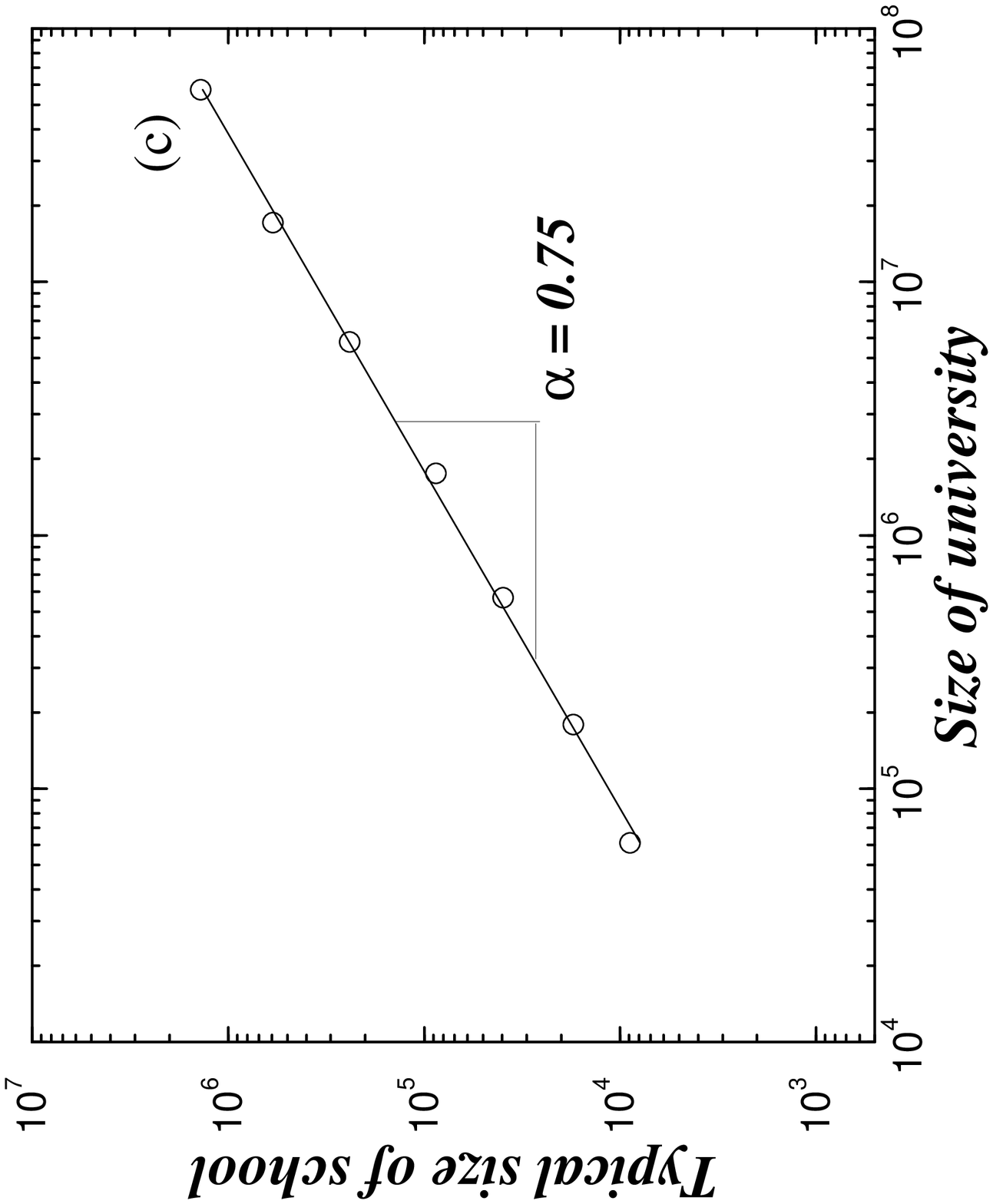}}}
}
\vspace*{0.5cm}
\centerline{
\epsfysize=0.8\columnwidth{\rotate[r]{\epsfbox{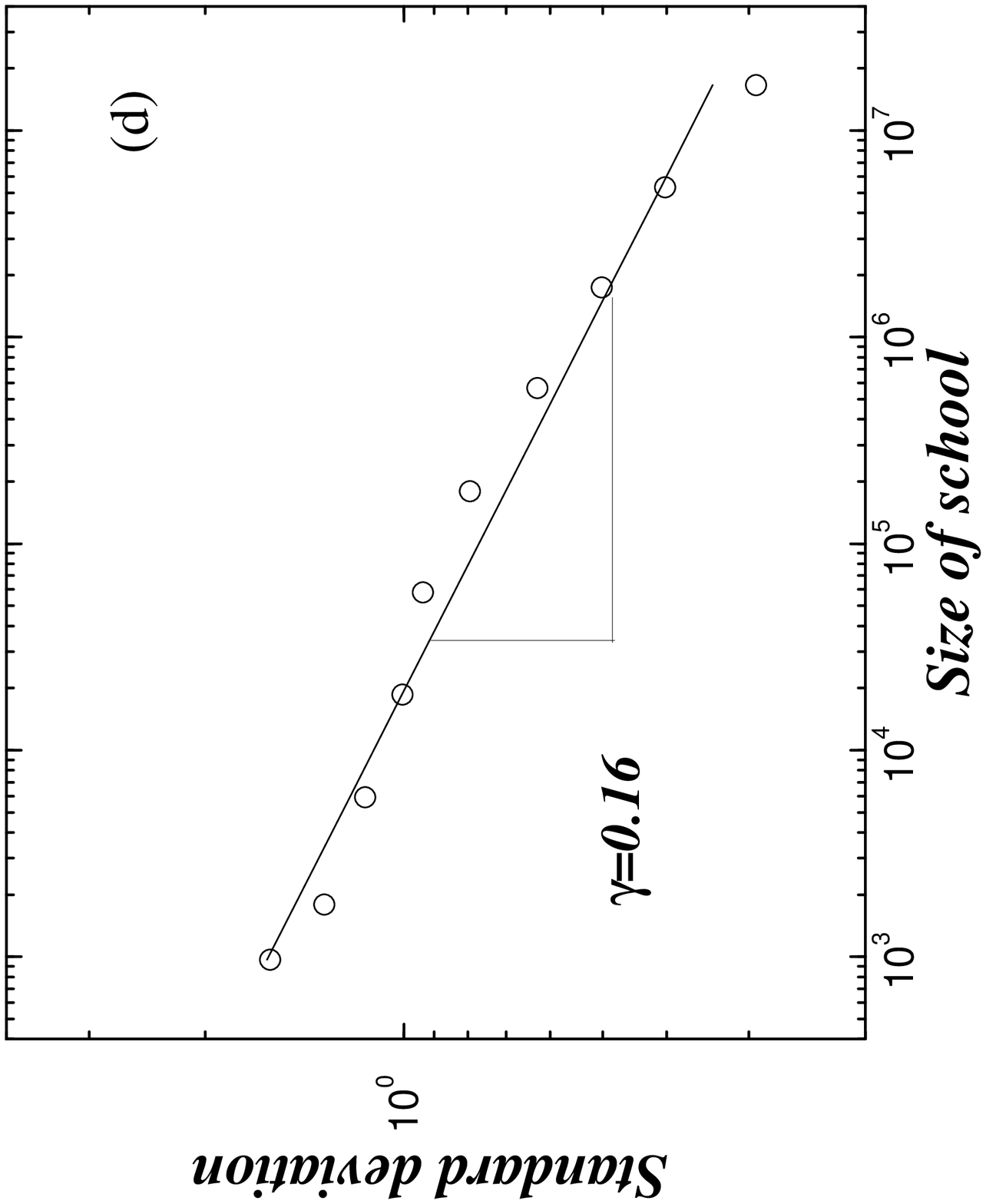}}}
}
\vspace*{0.5cm}
\caption{ Statistical analysis of the units forming the internal
structure of a university, the schools. {\it a)} Conditional
probability function $\rho(\xi|S)$ of finding a school of size $\xi$
in a university of size $S$.  To improve the statistics, we partition
the universities by size into four groups.  {\it b)} To illustrate the
scaling relation (\protect\ref{e-self}), we plot the scaled
probability density $\rho(\xi|S) / S^{-\alpha}$ versus the scaled size
of the school $\xi / S^{\alpha}$.  In agreement with
(\protect\ref{e-self}), we find that the scaled data fall onto a
single curve.  {\it c)} Scaling of the typical size of a school in a
university of a given size for different university sizes.  The data
obey a power law with exponent $\alpha = 0.75\pm0.05$.  {\it d)}
Standard deviation ${\cal W}$ of the distribution of growth rates of
schools versus school size $\xi$.  The data obey a power law with
exponent $\gamma = 0.16\pm0.05$.  Using (\protect\ref{e-beta}) and
this value of $\gamma$, we obtain an independent estimate $\beta =
0.25 \pm 0.05$. }
\label{f.self-similar}
\end{figure}

  Business firms are comprised of divisions and universities are made
up of schools or colleges, so it is natural to consider the internal
structure of these complex organizations\cite{Jovanovic93}.  We next
quantify how the internal structure of a university depends on its
size by calculating the conditional probability density $\rho(\xi|S)$
to find a school of size $\xi$ in a university of size $S$
(Fig.~\ref{f.self-similar}a).  The model predicts that $\rho(\xi|S)$
obeys the scaling form\cite{Amaral98}
\begin{equation}
\rho(\xi|S) \sim S^{-\alpha} f \left(\xi / S^{\alpha}
\right) \,,
\label{e-self}
\end{equation}
where $f(u) \sim u^{-\tau}$ for $u \ll 1$, and $f(u)$ decays as a
stretched exponential for $u \gg 1$.  We find $\tau = 0.37\pm 0.10$
(Fig.~\ref{f.self-similar}b), and $\alpha = 0.75\pm0.05$
(Fig.~\ref{f.self-similar}c). We test the scaling hypothesis
(\ref{e-self}) by plotting the scaled variables $\rho(\xi|S) /
S^{-\alpha}$ versus $\xi / S^{\alpha}$.  Figure~\ref{f.self-similar}b
shows that all curves collapse onto a single curve, which is the
scaling function $f(u)$.

  Equation~(\ref{e-self}) implies that the typical number of schools
with research activities in a university of size $S$ scales as
$S^{1-\alpha}$, while the typical size of these schools scales as
$S^{\alpha}$. Hence, we can calculate how $\sigma$ depends on $S$,
\begin{equation}
\sigma(S) \sim (S^{1-\alpha})^{-1/2} ~ {\cal W}(\xi)\,.
\label{e-width}
\end{equation}
In order to determine $\sigma$, we first find the dependence of ${\cal
W}$ on $\xi$.  Figure~\ref{f.self-similar}d shows that ${\cal W} \sim
\xi^{-\gamma}$ with $\gamma = 0.16\pm0.05$.  Substituting into
(\ref{e-width}) and remembering that the typical size of the schools
is $S^{\alpha}$, we obtain $\sigma(S) \sim (S^{1-\alpha})^{-1/2}
(S^{\alpha})^{-\gamma}\,$, which leads to the testable exponent
relation
\begin{equation}
\beta = \frac{1 - \alpha}{2} + \alpha \gamma\,.
\label{e-beta}
\end{equation}
For $\alpha \approx 3/4$ and $\gamma \approx 1/6$, Eq.~(\ref{e-beta})
predicts $\beta \approx 1/4$, in surprising agreement with our
empirical estimate of $\beta$ from the five distinct databases
analyzed (Fig.~\ref{f.collapse}a).


  Our results are consistent with the possibility that the statistical
properties of university research activities are surprisingly similar
for different measures of research activity and for distinct academic
systems.  Moreover, our findings for university research resemble
those independently found for business
firms\cite{Gibrat31,Simon77,Sutton97,Stanley96,Takayasu98} and
countries\cite{Summers91,Durlauf96,Lee98}. One possible explanation is
that peer review, together with government oversight, may lead to an
outcome similar to that induced by market forces, where the analog of
peer-review quality control may be consumer evaluation, and the analog
of government oversight may be product regulation.


ACKNOWLEDGEMENTS: We are indebted to the referees of this manuscript
for helpful suggestions which motivated additional analysis of four
databases reported here and the possible practical implications
discussed. We thank M.~Barth\'elemy, S.V.~Buldyrev, D.~Canning,
X.~Gabaix, S.~Havlin, P.Ch.~Ivanov, H.~Kallabis,~Y. Lee, and
B.~Roehner for stimulating discussions. We also thank N.~Bayers,
E.~Garfield, and especially R.E.~Hudson for help with obtaining the
ISI database.  We thank NSF, and LANA thanks FCT/Portugal for
financial support.


\vspace*{1cm}
\vfill
\eject

\end{multicols}
\end{document}